\documentclass[a4paper,twocolumn]{article}
\pdfoutput=1
\usepackage{fullpage}
\usepackage[inline]{enumitem}
\usepackage{amsmath}
\usepackage{amssymb}
\usepackage{pifont}
\usepackage{hyperref}
\usepackage{cite}
\usepackage{doi}
\usepackage{xargs}
\usepackage{rotating}
\usepackage{longtable}
\usepackage{booktabs}
\usepackage{multirow}
\usepackage{caption}

\newcommand{\sidetext}[1]{\begin{sideways}#1\end{sideways}}

\newcommand{\cmark}{\ding{51}}
\newcommand{\xmark}{\ding{55}}
\newcommand{\mytilde}{{\raise.17ex\hbox{$\scriptstyle\sim$}}}

\title{\textbf{Survey of Graph Analysis Applications}}
\author{
	Tim Hegeman\\
	Delft University of Technology\\
	\texttt{T.M.Hegeman@student.tudelft.nl}\\
	\and
	Alexandru Iosup\\
	Vrije Universiteit Amsterdam\\
	\texttt{A.Iosup@vu.nl}
}
\date{\today}

\begin{document}
\maketitle

\begin{abstract}
	Recently, many systems for graph analysis have been developed to address the growing needs of both industry and academia to study complex graphs.
	Insight into the practical uses of graph analysis will allow future developments of such systems to optimize for real-world usage, instead of targeting single use cases or hypothetical workloads.
	This insight may be derived from surveys on the applications of graph analysis.
	However, existing surveys are limited in the variety of application domains, datasets, and/or graph analysis techniques they study.
	In this work we present and apply a systematic method for identifying practical use cases of graph analysis.
	We identify commonly used graph features and analysis methods and use our findings to construct a taxonomy of graph analysis applications.
	We conclude that practical use cases of graph analysis cover a diverse set of graph features and analysis methods.
	Furthermore, most applications combine multiple features and methods.
	Our findings motivate further development of graph analysis systems to support a broader set of applications and to facilitate the combination of multiple analysis methods in an (interactive) workflow.
\end{abstract}

\section{Introduction}
\label{sec:intro}

Graph analysis is used across many application domains to interpret complex webs of relationships and connections formed by people, roads, financial transactions, etc. Understanding the practical uses of graph analysis is key to tuning existing graph analysis systems and guiding the development of new systems.
However, this understanding requires knowledge of applications from many domains, including the datasets and graph analysis methods they use.
Existing surveys focus on studying in-depth the datasets and analysis methods used in a single domain~\cite{bullmore2009complex,aittokallio2006graph,garcia2008some,peng2013survey,reps1998program}, identifying applications of specific (classes of) graph algorithms~\cite{fortunato2010community,conte2004thirty,demange2015some}, or exploring a variety of application domains~\cite{boccaletti2006complex,costa2011analyzing}.
In contrast, we identify applications across a large number of application domains and characterize the datasets and graph analysis methods used in practice.

To facilitate the growing need for analyzing graphs, many graph analysis systems have been developed\footnote{Doekemeijer et al.~\cite{doekemeijer2014survey} identified over 80 parallel graph processing frameworks between 2004 and 2014.}.
Most systems target generic applications of graph analysis, e.g., by providing a generic programming model like Pregel~\cite{malewicz2010pregel}, without explicitly considering the characteristics of real-world applications.
However, the performance of graph analysis applications depends a combination of three characteristics~\cite{guo2014well,iosup2016ldbc}, known as the PAD-triangle: the platform, the algorithm, and the dataset.
Thus, when developing and tuning graph analysis platforms, knowledge of the algorithms and datasets used in practice is essential to achieving good performance across many applications.

To address the gaps in knowledge left by previous work, we pose our main research question: \emph{What are the characteristics of the datasets and analysis methods used in practical applications of graph analysis?}
We further define two sub-questions: \emph{How to identify practical applications of graph analysis?} and \emph{How to characterize graph datasets and graph analysis methods?}
In this work we make three contributions to answer these questions:
\begin{enumerate}
	\item We present a systematic method for identifying practical use cases of graph analysis (Section~\ref{sec:method}) and apply this method to find a set of graph analysis application (Section~\ref{sec:applications}).
	\item We identify commonly used graph features (Section~\ref{sec:graph}) and classes of graph analysis methods (Section~\ref{sec:analysis}). Based on these common elements, we present a taxonomy of graph analysis applications (Section~\ref{sec:taxonomy}).
	\item We propose directions for future research in the development of graph analysis systems (Section~\ref{sec:future}).
\end{enumerate}

\section{Method for Finding, Selecting, and Characterizing Relevant Material}
\label{sec:method}

Applications of graph analysis can be found across many application domains and use a wide range of datasets and algorithms. In this section we define a method for finding and selecting literature on graph analysis applications and for characterizing their datasets and methods.

\subsection{Selection of a Comprehensive Method}
\label{sec:method:method}

\begin{table*}[!t]
	\centering
	\caption{Overview of steps that comprise the Systematic Literature Review method by Kitchenham et al.~\cite{kit_cha_2007} The steps we apply in this work are indicated by a checkmark (\cmark) and we list the section(s) implementing the step (if applicable).}
	\label{tab:method:kitchenham-steps}
	\begin{tabular}{cll}
		\toprule
		\multicolumn{3}{c}{\textbf{Planning the review}} \\
		\midrule
		\cmark & Identification of the need for a review & (S.~\ref{sec:intro}) \\ 
		\xmark & Commissioning a review & \\ 
		\cmark & Specifying the research question(s) & (S.~\ref{sec:intro}) \\ 
		\cmark & Developing a review protocol & (S.~\ref{sec:method:method}) \\ 
		\xmark & Evaluating the review protocol & \\ 
		\midrule
		\multicolumn{3}{c}{\textbf{Conducting the review}} \\
		\midrule 
		\cmark & Identification of research & (S.~\ref{sec:method:identification}) \\ 
		\cmark & Selection of primary studies & (S.~\ref{sec:method:selection}) \\ 
		\xmark & Study quality assessment & \\ 
		\midrule
		\multicolumn{3}{c}{\textbf{Reporting the review}} \\
		\midrule
		\cmark & Data extraction and monitoring & (S.~\ref{sec:method:analysis},~\ref{sec:applications},~\ref{sec:taxonomy}) \\ 
		\cmark & Data synthesis & (S.~\ref{sec:graph}-\ref{sec:taxonomy})\\ 
		\xmark & Specifying dissemination mechanisms & \\ 
		\cmark & Formatting the main report & \\ 
		\xmark & Evaluating the report & \\ 
		\bottomrule
	\end{tabular}
\end{table*}

Three common methods used to conduct literature surveys are: unguided traversal of the material, snowballing~\cite{DBLP:conf/ease/Wohlin14,DBLP:journals/misq/WebsterW02}, and the Systematic Literature Review method proposed by Kitchenham et al.~\cite{kit_cha_2007} Unguided traversal of the material is the simple process of reading as much as possible of the topic starting from a seed set of articles (e.g., provided by the supervisor) and continuing with as many articles as the reader can find using the typical repositories and search tools for scientific literature. For example, the reader could pursue every relevant link in each article read, or check all articles in the best conferences and journals in the past decade. The unguided element of the method comes from the lack of definition of stop and search criteria. The decision of which articles to select for review from the set of found articles is left entirely to reader and is not guided by a set of selection criteria. The snowballing method uses similar mechanisms, but imposes guidance elements such as criteria for finding and selecting material.

The Systematic Literature Review method of Kitchenham et al. is a comprehensive method for conducting literature reviews. As summarized in Table~\ref{tab:method:kitchenham-steps}, the methods consists of three major stages: planning, conducting, and reporting. Each of these stages is comprised of a set of steps whose application depends on the application domain and the specific goals of the survey. The stage of conducting the review in the SLR method includes at least three important elements: identifying the repositories and search engines that can deliver relevant material, defining a set of specific keywords (queries) used as automated selection criteria for relevant material, and defining a procedure for manually selecting truly relevant material from the set obtained through automated search.

We compare qualitatively these three methods, considering two criteria: the scale of the resulting dataset, and selection bias. The unguided traversal method can yield any amount of material, but it has an implicit selection bias towards material already known to the reader by relying on the reader to select directions for searching material. The snowballing method results in a large body of relevant material with a minor selection bias caused by the choice of a seed set of relevant material. The SLR method yields a limited set of relevant material and avoids a selection bias through a systematic search.

To make a selection, we first specify the desired outcomes for the two criteria. We prefer a limited number of articles because we would like to do an in-depth manual inspection of each, and we envision a large number of application domains (many of which we do not know) which requires a lack of selection bias. Therefor, we select the Systematic Literature Review method of Kitchenham et al. which best meets our criteria. In the remainder of this work, we follow the steps of the method as listed in Table~\ref{tab:method:kitchenham-steps} where applicable. Notably absent in our approach is the ``study quality assessment'' step. The quality of the work presented in surveyed material is largely irrelevant to our analysis; we consider only the datasets and graph analysis methods used and do not investigate how well these methods perform with respect to other approaches in a given domain.

\subsection{Identification and Selection of Literature}
\label{sec:method:identification}
\label{sec:method:selection}

\begin{table*}[!t]
	\centering
	\caption{Search queries used to identify relevant literature. $^\star$Three articles were retrieved via two search queries.}
	\label{tab:method:queries}
	\begin{tabular}{llcc}
		\toprule
		\textbf{ID} & \textbf{Query} & \textbf{\# Selected} & \textbf{\# Analyzed} \\
		\midrule
		Q1 & graph analysis & 39 & 34 \\ 
		Q2 & graph analytics & 4 & 3\\ 
		Q3 & graph mining & 19 & 14 \\ 
		Q4 & graph processing & 3 & 2 \\ 
		Q5 & network analysis & 15 & 7 \\ 
		Q6 & network analytics & 6 & 2 \\ 
		Q7 & network mining & 12 & 1 \\ 
		Q8 & network processing & 1 & 0 \\ 
		\midrule
		& \textbf{Total:} & $96^\star$ & $60^\star$ \\
		\bottomrule
	\end{tabular}
\end{table*}

To identify relevant literature we considered typical search engines for scientific literature, as recommended by the SLR method. Due to the wide range of application domains that potentially use graph analysis, we excluded any search engine dedicated to specific fields. We selected Google Scholar for its extensive corpus and open access.
We used eight queries to search for relevant material as listed in Table~\ref{tab:method:queries}. For each query we retrieved the first 100 search results for further inspection.
We conducted our search during January 2018.

Our search queries were formulated to match a wide range of possible applications, but they also match a large volume of irrelevant material. We used manual selection to extract all relevant literature from the body of search results.
We selected all articles that explicitly describe the use of one or more algorithms or methods for graph analysis as a primary contribution to address their research question(s).
We specifically exclude secondary studies, books, and theses because they typically present multiple applications and inclusion of these applications may have led to overrepresentation of the corresponding application domain.
We also exclude articles presenting a system or algorithm for graph analysis unless they target a specific application (domain).

The results of the identification and selection process are summarized in Table~\ref{tab:method:queries}. From 800 search results we selected 96 relevant articles (12\%). We further reduced this set to 60 articles for in-depth analysis through manual selection while preserving the diversity of application domains among selected articles.

\subsection{Analysis of Selected Material}
\label{sec:method:analysis}

To analyze the selected material we used a three-step process. The primary purpose of this process is characterizing the datasets and methods for graph analysis used in practice.
First, we scanned each selected article to summarize the application it describes (presented in Section~\ref{sec:applications}). We also identified any notable features of the dataset and the primary method of graph analysis used by each application.
Second, we derived a list of common (classes of) graph features (Section~\ref{sec:graph}) and graph analysis methods (Section~\ref{sec:analysis}) from the initial analysis performed in the first step.
Third, we extracted from each selected article the graph features and graph analysis methods they use. The resulting data was used to construct a taxonomy as presented in Section~\ref{sec:taxonomy}.

We manually classified the graph features present in each application. Due to a combination of frequently used terms (e.g., many articles refer to both ``directed'' and ``undirected'' graphs) and features not explicitly identified by the authors (e.g., vertex and edge properties, heterogeneity), keyword-based search as primary classification method was not feasible. Where plausible we used keyword-based search to validate the results of the manual inspection process (used keywords are listed in Section~\ref{sec:graph} where applicable).

To classify the graph analysis methods used by each application, we used search queries for most classes of methods (listed in Section~\ref{sec:method} where applicable), followed by manual inspection of the context of each search result to rule out false positives. We supplemented these results with the list of primary methods extracted during the initial scan of each article, especially for (classes of) methods without well-defined terminology (e.g., no suitable keywords for the ``graph mutation'' class were found due to the ubiquity of candidates such as ``reduction'', ``merge'', ``mutate''). Some atypical methods for graph analysis may have been missed in our classification process if such a method was not identified as the primary method of analysis of an article.

\subsection{Threats to Validity}

Although we selected our method to identify relevant graph analysis applications from a broad range of domains, the specifics of our search strategy introduces three potential biases.
First, we restrict our search to scientific literature, so we do not identify any commercial applications if their methods have not been published.
Second, we restrict our search to English literature, which may exclude applications that are not well-known in the English-speaking scientific community.
Third, some of our search queries show a strong correspondence with specific types or analysis, e.g., ``mining'' returns many application of pattern matching or subgraph isomorphism, whereas ``network analysis'' often occurs in the phrase ``social network analysis'', referring to a common set of graph analysis methods.

\section{Applications}
\label{sec:applications}

In this section we present the graph analysis applications we have selected and characterized using the method presented in Section~\ref{sec:method}. Applications are grouped by application domain.

\subsection{Biology}
\label{sec:applications:biology}

Biological networks are used to study the interactions of numerous biological entities, including proteins, genes, and organisms. We present in turn the biological applications we characterized.

\textbf{Protein-protein interaction networks:}
Li et al.~\cite{li2005interaction} propose an algorithm to identify protein complexes in protein-protein interaction networks. Their Local Clique Merging Algorithm (LCMA) iteratively identifies local cliques and merges them if they overlap significantly (i.e., are similar).

\textbf{Gene regulatory networks:}
Pinna et al.~\cite{10.1371/journal.pone.0012912} address the problem of deriving a gene regulatory network from observed gene expression levels. The authors construct a network of genes and inferred edges with weights signifying an initial estimate of the probability of an edge's existence. Edges with a probability below some threshold are removed from the network. Using strongly connected components, the authors identify unessential edges (feed-forward edges) to derive a final regulatory network.

\textbf{Metabolic networks:}
Koyt\"{u}rk et al.~\cite{koyuturk2004efficient} present an algorithm for identifying common patterns in graph-based representations of metabolic pathways. Their graphs contain one vertex for every unique enzyme in a pathway and edges for interaction between those enzymes. By mining the metabolic pathway graphs of multiple organisms for frequent subgraphs, the authors are able to identify sub-pathways common to many organisms.

\textbf{Tissue modeling:}
Bilgin et al.~\cite{bilgin2007cell,bilgin2010ecm} classify tissue samples by segmenting tissue images to identify cells, linking those cells in cell graphs, computing various graph theoretical measures for each cell graph, and applying machine learning techniques. They apply variations of this method to detect cancer in breast tissue~\cite{bilgin2007cell} and bone tissue~\cite{bilgin2010ecm}.

\textbf{Microbial communities:}
Barber{\'a}n et al.~\cite{barberan2012using} study the co-occurrence of microbes. Their network consists of vertices corresponding to microbes and edges corresponding to statistically significant correlation in occurrence across soil samples. Based on several network measures, e.g., average path length, and a comparison of the network's structure to a random graph, the authors find a distinct separation between two types of microbes: generalists and specialists.

\textbf{Protein assembly:}
A typical approach to identifying proteins involves collecting information on the peptide fragments (i.e., parts of a protein) in a sample and assembling these fragments to find proteins that could have been present in the sample. Zhang et al.~\cite{zhang2007proteomic} solve the assembly problem using a bipartite graph: observed peptides and potential proteins are mapped to vertices and an edge is added between a peptide and all proteins it is part of. To find likely candidates for proteins present in the sample, the authors reduce the resulting graph by merging vertices with identical connections, extract all connected components, and use a greedy set cover algorithm to find a set of proteins that cover all observed peptides.

\textbf{Other:}
Royer et al.~\cite{10.1371/journal.pcbi.1000108} propose the \emph{Power Graph Analysis} method for compressing biological networks. They identify three basic motifs found in many networks: stars, cliques, and bicliques. By iteratively identifying motifs and replacing subgraphs with \emph{power nodes}, the authors achieve compression rates of up to 85\% without losing information for a variety of biological networks.

\subsection{Neuroscience}
\label{sec:applications:neuroscience}

The human brain consists of an estimated 100 billion neurons and 1 quadrillion synapses. Although these neurons and synapses naturally comprise a brain network, this network is difficult to collect and analyze due to the small scale of the neurons and the large scale of the resulting network. Instead, brain networks used for practical neuroscience applications consist of brain regions and their communication paths.

Brain networks collected using fMRI are typically undirected and weighted, with each weight representing the level of communication between two brain regions. These weighted graphs are typically converted to unweighted graphs by dropping all edges with a weight below some threshold. The resulting graph can be analyzed for typical small-world properties: high average clustering coefficient and relatively low characteristic path length~\cite{watts1998collective}. Medical conditions that have been studied using this technique include Alzheimer's disease~\cite{10.1371/journal.pone.0013788,supekar2008network}, brain tumors~\cite{BARTOLOMEI20062039}, and traumatic brain injury~\cite{caeyenberghs2012graph}. Weighted networks can also be studied directly, as done by Stam et al.~\cite{stam2008graph} for Alzheimer's disease patients. The authors further study two damage models, \emph{Random Failure} and \emph{Targeted Attack}, through simulation and find that the Targeted Attack model best approximates the deterioration in brain connectivity observed in AD patients.

A directed brain network can be obtained from \emph{EEG} recordings of electrical activity in the cerebral cortex. This type of network has also been shown to have typical small-world properties~\cite{zamora2009graph}: high average clustering coefficient and low average path length. Various studies have shown deviations in connectivity when affected by certain medical conditions. For example, patients with epilepsy have more regular (i.e., not centralized) activity between brain regions during a seizure~\cite{wilke2011graph}. Spinal cord injured patients show larger levels of internal organization and fault tolerance, possibly as compensation triggered by the injury~\cite{fallani2007cortical}.

\subsection{Security}
\label{sec:applications:security}

Graph analysis powers a variety of security applications, most notably to identify vulnerabilities or anomalies in computer networks.

\textbf{Network vulnerability analysis:}
Phillips et al.~\cite{DBLP:conf/nspw/PhillipsS98} present a system for analyzing vulnerabilities in computer networks using attack graphs. They generate attack graphs from attack templates, i.e., specifications of the pre- and post-conditions of an attack. These templates are combined in a graph in which each node represents a combination of machines, users, and/or permissions that the attacker has obtained and each edge represents an action that may be performed by the attacker to compromise a new machine, obtain more permissions, etc. A cost or probability of success may be associated with each edge to study the most likely paths of intrusion. The authors further describe a variety of techniques to improve security using their system, including selecting from a set of possible security measures the most cost effective, or determining a minimal set of monitors to place such that each attack may be detected by multiple monitors.

Noel et al.~\cite{DBLP:conf/csiirw/NoelJ14} define a suite of metrics to quantify the vulnerability of a network based on its attack graph, including three metrics based on graph theoretical properties. First, they identify weakly connected components. The presence of separate components in the attack graph is indicative of a lack of vulnerabilities between multiple sets of machines in the network. Second, they identify cycles. Infecting any machine in a cycle gives indirect access to all other machines in the cycle, which represents a larger surface area for attacks. Third, they compute the diameter. A large diameter implies a large number of actions that an attacker needs to take to compromise the entire network.

\textbf{Malware detection:}
Kwon et al.~\cite{DBLP:conf/ccs/KwonMJBD15} analyze \emph{download graphs} to identify \emph{droppers}, malicious programs that download other programs (i.e., malware, or other droppers) to a host machine. The download graph consists of programs as vertices, and download relationships as edges. The authors extract subgraphs, called influence graphs, rooted in a single program vertex and containing all other programs that have been directly or transitively downloaded by the root program. The authors use a variety of metrics to summarize each influencer graph and use machine learning techniques to classify the root of each influence graph as a legitimate program or a dropper.

Polonium~\cite{chau2011polonium} detects malware by analyzing a large, bipartite graph of machines and files found on those machines. Polonium uses a belief propagation algorithm to estimate the probability of a file being malicious. This algorithm is seeded with external data on machine reputation and known malicious files.

\textbf{Botnet detection:}
Millions of computers worldwide have been infected by malicious software and are now part of networks of infected hosts called botnets. Many such botnets rely on peer-to-peer (P2P) communication to spread commands to members of the network. Nagaraja et al.~\cite{DBLP:conf/uss/NagarajaMHCB10} propose a method for identifying a P2P botnet in a communication graph, i.e., a graph in which every host is a vertex and two vertices are connected by an undirected edge if the corresponding hosts have communicated during some period of time. First, the authors use random walks on the communication graph to distinguish fast-mixing hosts (i.e., likely members of a P2P network) from slow-mixing hosts. Next, the vertices are clustered into subgraphs using k-means, and the nodes are clustered using an extension of SybilInfer [CITE], an algorithm based on random walks to extract a strongly connected group of nodes from a graph.

\textbf{Anomaly detection:}
Jiang et al.~\cite{DBLP:conf/icnp/JiangCJLZ10} analyze DNS traffic to identify anomalies. They model DNS failures as a bipartite graph mapping hosts that have issued at least one failed DNS request to domain names they have queried. Using a matrix factorization algorithm (tNMF), the authors decompose the DNS failure graph into communities of hosts and domains. They analyze each community and identify typical structures: stars and bi-meshes. Further, they track the evolution of communities over time.

\textbf{Video surveillance:}
Calderara et al.~\cite{DBLP:journals/cviu/CalderaraHPCT11} use spectral graph theory to analyze trajectories of people observed by video surveillance. The physical space observed in the video is quantized and the observed trajectories are translated to sequences of quantized locations. A graph is constructed with nodes representing a transition from one location to another and weighted edges representing the possibility of moving from one location to another location via precisely one intermediate location (shared by the source and target node). Spectral graph theory is used to filter out noisy trajectories from the graph and to determine if a new trajectory is either consistent with earlier observation or anomalous.

\subsection{Software Engineering}
\label{sec:applications:software-engineering}

Graphs occur naturally in various aspects of software development, e.g., call graphs, control flow graphs, dependency graphs. We describe several applications of graph analysis in software development and related activities.

\textbf{Identifying and locating software bugs:}
Cheng et al.~\cite{DBLP:conf/issta/ChengLZWY09} localize bugs by mining software behavior graphs. Execution traces of a program are captured as method-level or block-level behavior graphs in which each vertex presents a method or basic block, respectively. Edges in software behavior graphs can capture a variety of control flow relationships, e.g., one method calling another. The authors capture behavior graphs from multiple executions of the same program and label which behavior graphs correspond to faulty executions. Next, they mine the most discriminative subgraph(s) to localize the difference(s) in execution between correct and faulty behavior.

Maxwell et al.~\cite{DBLP:conf/kdd/MaxwellBR10} identify memory leaks in heap dumps by mining recurring patterns of object references. Their heap graph representation contains a vertex for every object in the heap dump and an edge for every reference. The authors first reduce the graph by extracting the dominator tree and using graph grammar reduction to compress typical (recursive) patterns. Next, they mine frequent subgraphs to identify potential memory leaks.

Eichinger et al.~\cite{DBLP:conf/pkdd/EichingerBH08} locate software bugs in call graphs using graph mining. Their approach first reduces call graphs by identifying sequences of identical calls and merging the corresponding subgraphs. Edge weights are introduced to capture the frequencies of calls. Next, the authors use weighted frequent subgraph mining to identify differences between successful and unsuccessful executions, thus revealing potential locations of bugs.

\textbf{Defect prediction:}
Zimmermann et al.~\cite{DBLP:conf/icse/ZimmermannN08} use network measures on software dependency graphs to identify critical binaries and predict defects. They find significant correlation between several network measures (e.g., eigenvector and degree centrality) and the number of defects in a binary. They find that network measures are able to predict 60\% of defects, compared to 30\% for traditional software complexity measures.

\textbf{Diagnosing distributed systems:}
G\textsuperscript{2}~\cite{DBLP:conf/usenix/GuoZLYLDLZ11} is a graph processing system for analyzing large software execution graphs, i.e., graphs describing system events and their relationships. G\textsuperscript{2} can be programmed to perform custom queries on an execution graph. Underlying these queries are two key operations: \emph{slicing} (i.e., extracting all events that are causally related to a queried event) and \emph{hierarchical aggregation} (i.e., merging a set of related events into a single event).

\textbf{Software plagiarism detection:}
GPLAG~\cite{DBLP:conf/kdd/LiuCHY06} detects software plagiarism by representing programs as program dependence graphs. Statements in the program are encoded as vertices, and control flow and data dependencies are encoded as directed edges. Type information is encoded as vertex and edge properties. Commonalities between two program are detected using subgraph isomorphism.

\textbf{Developer collaboration:}
Surian et al.~\cite{DBLP:conf/wcre/SurianLL10} analyze the SourceForge collaboration network to discover how well connected developers are, what topological structures characterize developer communities, etc. The authors first identify connected components and mark each as either a small or a large community. Second, they mine common patterns among the small communities. Third, they compute the frequency of each pattern in both the small and large communities.

\subsection{Logistics \& Planning}
\label{sec:applications:logistics-planning}

Navigation systems are a stereotypical example of real-world graph analysis applications. More generally, logistics and planning problems are often solved using graphs.

\textbf{Road networks:}
Urban street networks can be naturally modeled as a graph; intersections can be mapped to vertices and the roads connecting intersections can be mapped to edges. This is typically referred to as the \emph{primal} graph representation. For some applications an alternative representation may be preferred. The \emph{dual} graph representation maps roads to vertices and intersections to edges. Porta et al. describe various methods for analyzing both the primal~\cite{porta2006networkPrimal} and dual~\cite{porta2006networkDual} urban street graphs.

TrajGraph~\cite{DBLP:journals/tvcg/HuangZMYYZ16} uses graph-based visual analytics to study traffic in urban street networks, in particular using taxi route data. It uses partitioning techniques to reduce the size of the graph for visualization purposes. Further, TrajGraph provides automated analysis of traffic to identify hubs at different points in time. Finally, it allows users to highlight arbitrary areas (i.e., subgraphs) to analyze local traffic.

\textbf{General planning problems:}
GraphPlan~\cite{DBLP:journals/ai/BlumF97} represents a planning problem as a \emph{Planning Graph}, a leveled DAG in which nodes represent propositions (states?), edges represent action, and levels represent timesteps in the produced plan. The graph is dynamically constructed and pruned one level at a time. A custom heuristic-based traversal algorithm is used to find valid plans.

Hong et al.~\cite{DBLP:journals/jair/Hong01} propose a \emph{goal graph}-based approach to goal recognition, loosely based on GraphPlan. They use an iterative two-stage algorithm to repeatedly extend the goal graph based on observed actions, followed by the identification of plans and goals that are consistent with the new graph.

Helmert~\cite{DBLP:conf/aips/Helmert04} proposes a greedy algorithm based on a causal graph for planning problems. The causal graph encodes state variables as vertices and possible causal relationships as directed edges (i.e., an edge from vertex A to vertex B indicates that the value of state variable B may depend on the value of state variable A). Further, every state variable has an associated domain transition graph with a vertex for every possible value of the variable and a directed edge for every transition, annotated with the preconditions for the transition. A plan is found using a heuristic-based traversal of the causal graph. The heuristic includes repeated shortest path searches on domain transition graphs.

\textbf{Network routing:}
Daly et al.~\cite{DBLP:conf/mobihoc/DalyH07} apply social network analysis to routing in mobile ad hoc networks (MANETs). Routing in MANETs is challenging because the network graph is rarely connected. Efficiently delivering messages requires identifying devices that are likely to connect to many other devices, thus quickly spreading the message through the network. By using network analysis techniques, the authors identify devices in the network with high betweenness (i.e., short routes to many other devices) as key candidates for spreading messages.

\subsection{Social Sciences}
\label{sec:applications:social-sciences}

Communication between people has long been studied to understand how groups of people interact. One-on-one communication methods, including (e-)mail, phone calls, and text messaging, can be modeled as a social interaction network in which each vertex represent a person and edges represent that two people have communicated. Similarly, social media platforms like Facebook and Twitter naturally comprise a social interaction network, often with additional information on types of interactions (friendships, follower-relationships, likes, etc.).

Schwartz et al.~\cite{DBLP:journals/cacm/SchwartzW93} analyze a directed interaction graph derived from mail traffic to discover shared interests. They apply typical SNA techniques, including extracting the largest weakly connected component, computing the diameter, average path length, etc. Further, they cluster the graph using \emph{Aggregate Specialization Graph Isolation} and \emph{Specialization Subgraph Derivation} to identify communities with shared interests.

Recently, many studies have analyzed the Twitter graph to identify important users (\emph{influencers}), authoritative users, etc. For example, TURank~\cite{DBLP:conf/wise/YamaguchiTAK10} ranks Twitter users based on authority scores. These scores are computed using a variation of PageRank, called ObjectRank, on the \emph{user-tweet graph} consisting of users, tweets, follow relationships, post relationships, and retweet relationships. Khrabov et al.~\cite{DBLP:conf/socialcom/KhrabrovC10} identify influential Twitter users using a combination of PageRank and a custom ranking based on a user's number of mentions. By using dynamic metrics, they identify key users by periods of consecutive, accelerated growth in influence. Yang et al.~\cite{DBLP:conf/sigir/YangLLR12} analyze retweets using a variation of the HITS algorithm to identify interesting posts, i.e., posts that may be of interest to a wider audience than the direct neighborhood of the user who posted it. They first identify authoritative users in the user-retweeted-user graph, and next identify interesting tweets based on the authority of their creator and retweets. They conclude that, by considering both users and retweets, their method performs better than consider only retweets.

\subsection{Psychology}

Mota et al.~\cite{mota2014graph} analyze speech graphs derived from dream reports produced by schizophrenic, bipolar, and control subjects. A speech graph consists of a node for every unique word in a body of text and a directed edge for every consecutive pair of words. By comparing 14 attributes (e.g., largest connected component, repeated and parallel edges, cycles, clustering coefficient) of each speech graph, the authors achieve high accuracy in classifying the group of a subject based on their dream report.

Verbal fluency tests are used to assess a person's ability to produce a sequence of words satisfying some task. For example, in a category fluency test, the subject is asked to list as many words fitting the given category as they can in a limited time. By converting this sequence of words to a speech graph, the subject speech patterns can be analyzed. Lerner et al.~\cite{lerner2009network} study the results of category fluency tests taken by adults with mild cognitive impairment (MCI), with Alzheimer's disease (AD), or neither. They merged the speech graphs of all subjects in the same subject group. Analysis of the three resulting graphs, one for each subject group, reveals that they have typical small-world properties, but that those properties are less prevalent for the MCI and AD groups. This is consistent with earlier research. Bertola et al.~\cite{bertola2014graph} perform a similar study, but analyze each subject's speech graph individually. Using an extensive set of graph measures, they can accurately classify a subject to be in one of the three subject groups.

\subsection{Science}
\label{sec:applications:science}

By virtue of the incremental nature of scientific progress, publications naturally form networks through the citations that connect them. Such citation networks have been studied for many scientific communities. Jacovi et al.~\cite{DBLP:conf/cscw/JacoviSGUSM06} apply social network analytics techniques to study the citation graph of the CSCW conferences and related work. They identify communities and track their evolution over time. Further, they identify chasm-papers; papers that were influential outside the CSCW conferences, but overlooked in the CSCW community. Gondal~\cite{DBLP:journals/socnet/Gondal11} analyzes a citation network for an emergent research field. In addition to small-world properties, the author considers the presence of various patterns, including stars, papers sharing a large number of cited authors, and papers citing primarily authors from a single country. Tsatsaronis et al.~\cite{DBLP:conf/ercimdl/TsatsaronisVTRNSZ11} apply Power Graph Analysis, a method originally developed to analyze and visualize biological networks, to study the DBLP citation network and its evolution.

\subsection{Chemistry}
\label{sec:applications:chemistry}

Molecules can be represented by graphs with atoms as vertices and the bonds connecting atoms as edges. Large databases of such molecular graphs and the properties of the associated molecules have been compiled and are used in, e.g., pharmaceutical research to identify (fragments of) molecules with desirable properties for a new drug. Nijssen et al.~\cite{nijssen2004frequent} propose a method based on frequent subgraph mining to extract common molecule fragments from a set of molecules. This method may be used to identify molecule fragments that characterize a set of input molecules sharing a desirable property. Wegner er al.~\cite{wegner2006data} propose a method for classifying molecules by mining their corresponding molecular graphs for maximum common substructures.

\subsection{Finance}
\label{sec:applications:finance}

Iori et al.~\cite{iori2008network} study overnight inter-bank traffic of Italian banks between 1999 and 2002. They define a temporal graph in which each vertex corresponds to a bank and each edge indicates that at least one transfer occurred between two banks during the selected time period. Each edge is further annotated with the number and total volume of transfers. After analyzing various network measures, the authors conclude that while some microstructure characteristics were found (e.g., degree correlates strongly with the size of the bank), the network is somewhat random, which is indicative of an efficient system.
Wang et al.~\cite{wang2012visibility} use a different approach to study financial data as time series. They apply visibility graph analysis to study four time series related to China's quarterly economic growth between 1992 and 2010. They identify small-world properties in all corresponding graphs.

\subsection{Linguistics}
\label{sec:applications:linguistics}

Jiang et al.~\cite{DBLP:conf/sgai/JiangCSZ09} propose a graph-based approach to document classification. Each document is represented by a graph containing a variety of vertex types (e.g., word, part of speech) and edge types (e.g., part of speech to a word, word order). The authors use frequent subgraph mining on document sets to group documents that share similar, uncommon phrasing.
Biemann~\cite{biemann2006chinese} proposes Chinese Whispers, a graph clustering algorithm designed to address various natural language processing problems. Example applications studied with Chinese Whispers include language separation (i.e., detecting languages in word co-occurrence graphs) and word class acquisition (i.e., clustering a word co-occurrence graph to detect classes of words).

\subsection{Other}
\label{sec:applications:other}

We identified several other applications of graph analysis that do not fit any of the previously discussed application domains.

\textbf{Cardiographs:}
Jiang et al.~\cite{jiang2013visibility} study the impact of meditation on heartbeat dynamics using visibility graph analysis. They monitor the heart rate of volunteers before and during mediation. The authors find that meditation causes significant changes in heartbeat rhythms.

\textbf{Geometric constraint problem:}
Geometric constraint problems describe sets of geometric entities (e.g., points, lines) and constraints between pairs of entities (e.g., distance, angle) with the purpose of assigning to each object a position, orientation, etc. such that all constraints are met. Lee et al.~\cite{LEE2003103} propose an approach based on graph reduction. Their algorithm initially constructs a graph containing each geometric entity as a vertex and each constraint as an edge. Next, it iteratively identifies clusters based on \emph{degrees of freedom} analysis (typically small cycles) and collapses them to a single pseudo-geometric entity until the graph is reduced to a single vertex.

\textbf{Knowledge-based systems:}
Rodriguez~\cite{DBLP:journals/corr/abs-0903-0194} analyzes the relationships between datasets maintained by the Linked Data community. They create a graph in which every vertex represents a dataset and every directed edge indicates a reference from one dataset to another. They identify strongly connected communities that correspond well with known domains (e.g., biology, computer science).

\textbf{Recommendation systems:}
Mirza et al.~\cite{DBLP:journals/jiis/MirzaKR03} propose a method for evaluating recommendation systems by interpreting their outcomes as a pair of graphs: a social graph comprised of all users and a bidirectional edge for every pair of similar users, and a recommender graph which extends the social graph with all recommended artifacts. By analyzing the connected components in the recommender graph and comparing average path lengths from users to artifacts between the recommender graph and a random graph, the authors evaluate the effectiveness of a recommendation algorithm.

\textbf{Seismology:}
The visibility graph analysis method has been applied to seismology data to varying success. For example, analyzing earthquakes in Italy between 2005-2010 reveals a long-range correlation in their magnitudes~\cite{0295-5075-97-5-50002}. However, time-clustering was not detected by the method.

\textbf{Video analysis:}
Yeung et al.~\cite{DBLP:journals/cviu/YeungYL98} use a \emph{Scene Transition Graph} to represent shots (as vertices) and the transitions between them (as directed edges). The authors identify strongly connected components as scenes in the analyzed video.

\textbf{Web usage mining:}
Website operators track the browsing patterns of visitors to study how visitors interact with the website, and consequently to identify changes that can be made to the website to improve the browsing experience. Heydari et al.~\cite{heydari2009graph} propose a graph-based approach to web usage mining. They capture each visitor's traversal through a website as a graph by representing web pages as vertices and links followed from one page to another as edges. Vertices are weighted by the time a user spent on the corresponding page. The authors mine a set of browsing graphs to identify common browsing habits.

\section{Graph Features}
\label{sec:graph}

A \emph{graph} is a structure consisting of a set of \emph{vertices} and a set of vertex pairs (\emph{edges}). Graphs are typically used to represent a collection of entities and their pairwise relationships. Many applications extend this model to encode additional information: the strength of a relationship, specific properties of an entity, differentiation of relationships types, etc. In this section we identify several commonly used \emph{graph features} among the surveyed applications presented in Section~\ref{sec:applications}. We present typical use cases for each feature.

\subsection{Edge Orientation}
\label{sec:graph:orientation}

Edges in a graph may be either directed, indicating a one-way relationship between two vertices, or undirected, indicating a two-way relationship. The types of relationships used by the surveyed applications are diverse and we did not identify a clear classification of use cases of either directed or undirected edges. However, we highlight several recurring relationship types and other observations. First, directed edges are frequently used in applications that analyze sequential or causally-connected processes, such as tracing movement through time~\cite{DBLP:journals/cviu/CalderaraHPCT11}, analyzing software call graphs~\cite{DBLP:conf/pkdd/EichingerBH08}, finding valid plans in a planning graph~\cite{DBLP:journals/ai/BlumF97}, or capturing a person's speech to understand psychological disorders~\cite{bertola2014graph}. Directed edges are also prevalent in sociology graphs to represent social interactions that are typically initiated by one party. For some types of graphs we found both directed and undirected variations, e.g., some methods for measuring brain activity do not capture the direction of communication, but other methods do.

\subsection{Weighted Graphs}
\label{sec:graph:weighted}

\emph{Keywords used to validate manual inspection:  weight, length, distance, strength.}

A typical extension of the graph model is the addition of weights to edges (and, less commonly, to vertices). That is, the definition of a graph is extended with a function that maps every edge (or vertex) to a single, typically real-valued, weight. Although weights may be used to model any property, we highlight two typical use cases of edge weights.

First, edge weights are commonly used to represent the strength of a relationship. For example, the edges in a brain network model communication paths between two regions. Their edge weights model the amount of activity, e.g., in~\cite{10.1371/journal.pone.0013788,BARTOLOMEI20062039,wilke2011graph,fallani2007cortical,caeyenberghs2012graph,stam2008graph,supekar2008network}. Similarly, edges can model the frequency of a relationship's occurrence, e.g., the number of calls made to a function~\cite{DBLP:conf/pkdd/EichingerBH08}, the number of co-authored papers~\cite{DBLP:conf/ercimdl/TsatsaronisVTRNSZ11}, or the number of bank transfers~\cite{iori2008network}.

Second, edge weights may be used to represent the length or distance of a connection. The stereotypical example for this use of edge weights is road networks. In such a network, edges often represent roads and their weights represent the length of roads (or the expected time taken to travel from the start to the end of the road). This application has been described in~\cite{porta2006networkPrimal} and many introductory text books on graph theory.

Only the first use case of edge weights as a measure of relationship strength was prevalent among the application presented in Section~\ref{sec:applications}. However, shortest path queries, an important class of graph analysis methods described in Section~\ref{sec:analysis:paths}, operate on either unweighted graphs or on weighted graphs representing distances. To accommodate shortest path queries on graphs with edge strengths, two approaches are commonly used. First, the weighted graph may be converted to an unweighted graph by removing all edges with a strength below some threshold and removing the weights of all remaining edges. This approach is frequently used for brain networks~\cite{10.1371/journal.pone.0013788,BARTOLOMEI20062039,wilke2011graph,fallani2007cortical,caeyenberghs2012graph,supekar2008network}. Second, the edge strengths may be converted to lengths by defining the length of an edge to be the inverse of its strength. Thus, strong edges are short and a path along several strong relationships may be shorter than a direct path via one weak relationship. This approach has been applied to citation networks~\cite{newman2001scientific}, brain networks~\cite{rubinov2010complex}, etc.

\subsection{Heterogeneous Graphs}
\label{sec:graph:heterogeneous}

By definition, graphs consist of one set of vertices and one set of edges. In some graphs, all vertices and edges are homogeneous. That is, they represent instances of one type of entity or relationship, respectively. However, some applications include multiple types of entities and relationship which are combined in a single heterogeneous graph. These types may be encoded in the graph representation as vertex and edge labels\footnote{In graph theory, labeling most commonly refers to assigning a unique label (or identifier) to every vertex/edge, typically integers from the range $[1, \text{\#vertices}]$ or $[1, \text{\#edges}]$. More generally, a vertex or edge labeling function assigns a label to every vertex or edge, respectively, without restriction.}. To avoid confusion with other uses of graph labels, we use the terms \emph{vertex/entity type} and \emph{edge/relationship type} to distinguish types of entities and relationships modeled by a heterogeneous graph.

Heterogeneity can have significant impact on the structure of a graph. For example, some types of relationships may only occur between two particular types of entities, some types of entities may not be able to have direct relationships with other types, and some types of relationships may be more numerous than others. Each of these examples introduces a bias in the number and location of relationships in a graph that can not be explained by topological measures such as degree distribution.

Examples of heterogeneity in vertex types include user and tweet vertices in a Twitter graph~\cite{DBLP:conf/wise/YamaguchiTAK10}, hosts and domain names in a DNS network~\cite{DBLP:conf/icnp/JiangCJLZ10}, and proteins and peptides in proteomics~\cite{zhang2007proteomic}. Examples of heterogeneity in edge types include dependencies between program statement in a call graph~\cite{DBLP:conf/kdd/LiuCHY06}, causal relationships in diagnosing distributed systems~\cite{DBLP:conf/usenix/GuoZLYLDLZ11}, and relationships between words in text classification~\cite{DBLP:conf/sgai/JiangCSZ09}.

\subsection{Property Graphs}
\label{sec:graph:property}

Weighted and heterogeneous graphs both extend graphs with a function mapping vertices and/or edges to a single weight or type. Property graphs generalize this approach to an arbitrary number of functions mapping to arbitrary values. Each function represents a property of the modeled entity or relationship. For example, in a social network one function maps each user vertex to its name and another function maps each friendship edge to the date the friendship was formed.

In our characterization, we use a more restricted definition of property graphs; an application is characterized as using property graphs if and only if the analysis of the graph uses the values of these properties. Although the entities and relationships of almost all applications have properties, these properties are not typically used in the analysis of the corresponding graph. For example, a social network application may identify communities by considering only the edges present in the graph and not considering the name, date of birth, or other personal information of each user. This application can use any graph analysis system without support for property graphs. Thus, we do not consider such an application to have a property graph.

Examples of property graphs include chemical structure graphs annotated with atom and bond characteristics~\cite{wegner2006data,nijssen2004frequent}, and an attack graph annotated with pre- and post-conditions for modeling attack vectors in a compute network~\cite{DBLP:conf/nspw/PhillipsS98}.

\subsection{Temporal Graphs}

\emph{Keywords used to validate manual inspection:  temporal, timestamp, evolution.}

In many real-world applications, graphs evolve over time as the entities and relationships captured in these graphs change. Most surveyed applications do not address this evolution in their analysis; the graphs analyzed by these applications represent snapshots of the real-world networks they model. In contrast, some applications use the temporal nature of their graphs explicitly, e.g., to study the evolution of the network. Graphs that include temporal data (e.g., the time at which a relationship was established) are known as temporal graphs\footnote{In our taxonomy we classify an application as using temporal graphs if and only if the application explicitly uses timestamped data for analysis. That is, even if a graph consists of timestamped data or data collected over a period of time, it is not classified as a temporal graph if the temporal data is not explicitly analyzed.}. 

We identify three use cases for temporal graphs. First, they are used to study the evolution of relationships in networks. For example, a citation graph annotated with the publication date of each paper can reveal how scientific communities interact and co-evolve~\cite{DBLP:conf/cscw/JacoviSGUSM06}. Second, some temporal graphs have static structures but dynamic weights or other properties, e.g., traffic data per hour in a road network~\cite{DBLP:journals/tvcg/HuangZMYYZ16}, or brain activity levels captured periodically~\cite{wilke2011graph}. Third, temporal graphs may be used to study sequences of events connected by causality relationships, e.g., communication via messages in a distributed systems~\cite{DBLP:conf/usenix/GuoZLYLDLZ11}.

\section{Methods for Graph Analysis}
\label{sec:analysis}

Many applications that operate on graphs use graph analysis techniques to extract information from a graph that is not evident from individual entities or relationships. For example, the degree of a single vertex in a social network graph may inform us how many friends the corresponding user has, but by analyzing all relationships in the graph, we can determine whether the user is popular, how important they are for cohesion in the network, what communities they are part of, etc. In this section we characterize the methods for graph analysis used by the applications presented in Section~\ref{sec:applications}.

\subsection{Neighborhood Statistics}
\label{sec:analysis:neighborhood}

\emph{Keywords used to identify uses: clustering coefficient, degree distribution, degree assortativity, local efficiency, local subgraph.}

Many graph-based applications use metrics to quantify a graph's structural properties. We refer to a common class of such metrics as \emph{neighborhood statistics}. These statistics are computed for each vertex in a graph and depend only on the immediate neighborhood of a vertex\footnote{The neighborhood of a vertex includes the vertex itself, all vertices with an edge to or from the target vertex, and all edges that connect two vertices included in the neighborhood}.

Two common neighborhood statistics are the degree distribution and local clustering coefficient of graph. The degree distribution of a graph is used across many application domains (e.g., biology~\cite{fallani2007cortical}, finance~\cite{wang2012visibility}) to support the hypothesis that a graph's structure is different from a random graph\footnote{Studies across many domains have found that their graphs are power-law graphs, i.e., that the degree distribution of the graph follows a power-law distribution. This matches the observations of Barab{\'a}si and Albert~\cite{barabasi1999emergence} who predicted the presence of power-law distribution across many large complex networks. However, the ``power-law hypothesis'' has frequently been argued against, including in a recent study by Broido and Clauset~\cite{DBLP:journals/corr/abs-1801-03400} which found strong evidence of power-law properties in only 4\% of graphs they analyzed.}. Similarly, the local clustering coefficient is commonly used to characterize a graph as ``small-world''~\cite{watts1998collective}, i.e., a graph consisting of clusters of highly connected vertices and few interconnections between clusters.

Less commonly used statistics include local efficiency~\cite{latora2001efficient} as a measure of efficient communication among a vertex's neighbors (e.g., in brain networks~\cite{caeyenberghs2012graph,fallani2007cortical}, road networks~\cite{porta2006networkDual}) and degree assortativity~\cite{newman2002assortative} as a measure of how likely high-degree vertices are to be connected (e.g., in~\cite{wang2012visibility,DBLP:journals/corr/abs-0903-0194,porta2006networkDual,iori2008network}).

The performance of neighborhood analysis is often sensitive to the structure of the input graph. Graph analysis systems can benefit from exploiting data locality if a vertex's neighborhood is small. However, the presence of several highly connected vertices (e.g., as found in power-law graphs) can lead to a significant imbalance in computation required per vertex.

\subsection{Paths and Traversals}
\label{sec:analysis:paths}

\emph{Keywords used to identify uses: path length, shortest path, breadth-first, depth-first, traversal, global efficiency, random walk.}

Many applications of graph analysis are not restricted to the neighborhood of vertices, but instead involve the analysis of paths between vertices. Of particular interest are shortest path queries, i.e., identify a path of minimal total length between given two vertices (or minimal number of edges in unweighted graphs). Although these queries can be answered individually using Dijkstra's algorithm (or breadth-first search on unweighted graphs), many applications require computing the shortest path between multiple or all vertex pairs (e.g., using the Floyd-Warshall algorithm).

We identify three primary uses of shortest path algorithms. First, shortest paths can be used directly, e.g., to find an attack path with the highest probability of success in a network attack graph~\cite{DBLP:conf/nspw/PhillipsS98}. Second, the average path length can be used to identify ``small-world'' properties~\cite{watts1998collective} in a graph. Global efficiency~\cite{latora2001efficient}, the inverse of the average path length, is used to determine the efficiency of communication between brain regions~\cite{fallani2007cortical}, to identify potential software defects~\cite{DBLP:conf/icse/ZimmermannN08}, etc. Third, the maximum shortest path length (diameter) and related metrics (e.g., eccentricity) are used across many domains to study worst-case paths.

We further identify two applications of path-based graph analysis. First, the enumeration of near-shortest paths (i.e., listing all paths between two vertices that are no longer than $\rho$ times the shortest path for a given parameter $\rho$) is used to identify a broad set of security vulnerabilities in a network~\cite{DBLP:conf/nspw/PhillipsS98}. Second, the computation of number of shortest paths between two vertices (i.e., the path multiplicity of a vertex pair) is computed as a measure of flexibility in communication between brain regions~\cite{zamora2009graph}.

Closely related to path queries are graph traversals, i.e., traversing through a graph from a source vertex via its outgoing edges. Typical examples of traversals include breadth-first search and depth-first search, which are both commonly used as building blocks in more complex methods of analysis. In the applications presented in Section~\ref{sec:applications} we further identified the use of random walks to identify sub-networks of bots in a larger network~\cite{DBLP:conf/uss/NagarajaMHCB10}, and domain-specific traversals used for planning~\cite{DBLP:journals/ai/BlumF97,DBLP:conf/aips/Helmert04,DBLP:journals/jair/Hong01} and debugging distributed systems~\cite{DBLP:conf/usenix/GuoZLYLDLZ11}.

\subsection{Connectivity}
\label{sec:analysis:connectivity}

\emph{Keywords used to identify uses: connected components.}

Instead of identifying \emph{how} vertices in a graph are connected (i.e., discovering paths), some graph problems are concerned with determining \emph{if} vertices are connected. The typical application of analyzing the connectivity of a graph is identifying its connected components. That is, dividing a graph into components such that two vertices belong to the same component if and only if a path exists between them. For directed graphs we distinguish two types of connected components: weak and strong. In a weakly connected component, any pair of vertices is connected by an undirected path (i.e., a path that treats every directed edge as though it is undirected). In a strongly connected component, directed paths exist from each vertex to every other vertex in the component. We identify several use cases of identifying connected components in the reviewed articles.

First, some applications decompose their graph into connected components to analyze components individually. For example, the largest component(s) in a communication network~\cite{DBLP:journals/cacm/SchwartzW93} or DNS failure graph~\cite{DBLP:conf/icnp/JiangCJLZ10} contain the vast majority of interesting vertices. By extracting only those components, any further analysis is not biased by a large number of isolated vertices. For other applications all components are of interest, e.g., a matching problem in a bipartite protein-peptide graph can be split into one smaller sub-problem per connected component~\cite{zhang2007proteomic}.

Second, the number of components or size of the largest component can be used as metrics to classify graphs. For example, the fraction of vertices that belong to the largest connected component is a highly discriminative feature in classifying breast~\cite{bilgin2007cell} or bone~\cite{bilgin2010ecm} tissue samples for cancer diagnosis, or in classifying speech graphs~\cite{bertola2014graph,mota2014graph}. 

Third, a graph's components can be interpreted as communities or otherwise related entities. For example, strongly connected components in a scene transition graph correspond to story units~\cite{DBLP:journals/cviu/YeungYL98} and components in a network attack graph correspond to sets of servers with vulnerable interconnections~\cite{DBLP:conf/csiirw/NoelJ14}. Other methods for detecting communities in a graph are presented in Section~\ref{sec:analysis:centrality}.

\subsection{Centrality and Ranking}
\label{sec:analysis:centrality}

\emph{Keywords used to identify uses: centrality, PageRank, ranking, HITS.}

In most real-world networks, not all entities are of equal importance to the network. Key to numerous applications is identifying the most important entities in a network. This is typically achieved by assigning to each entity a score to signify its importance and deriving a ranking from these scores.

Most commonly-used measures of importance are based on vertex- or edge-centrality. The centrality of a vertex or edge is a measure of how important it is to the structure of a network. Three commonly used categories of centrality are betweenness centrality, closeness centrality, and degree centrality, as first popularized by Freeman~\cite{freeman1978centrality} in the context of social networks\footnote{Although Freeman's centrality measures were defined in earlier work by different authors, Freeman is often credited for providing an intuitive conceptual interpretation of centrality in the context of social networks and for identifying three types of centrality that cover important structural characteristics of social networks.}. More recently, PageRank centrality~\cite{page1999pagerank} has seen adoption across application domains after first being used to identify important websites in the Google search engine.

Other measures for ranking used by reviewed applications include: information centrality in software dependency graphs~\cite{DBLP:conf/icse/ZimmermannN08}, straightness centrality in road networks~\cite{porta2006networkPrimal}, ObjectRank as adaptation of PageRank to a heterogeneous Twitter graph~\cite{DBLP:conf/wise/YamaguchiTAK10}, and Hyperlink-Induced Topic Search (HITS) to identify hubs and authorities (e.g., on Twitter~\cite{DBLP:conf/sigir/YangLLR12}).

\subsection{Clustering}
\label{sec:analysis:clustering}

\emph{Keywords used to identify uses: cluster(ing), community/communities.}

Clustering is the act of identifying clusters (or groups) of entities in a graph. Clustering techniques are often used to extract groups of related entities from a graph based on the relationships they have formed. Because the definition of ``related entities'' is application-specific, there exist many specialized algorithms for clustering. Among the surveyed applications, we identify two main approaches. First, some applications decompose a graph into well-defined groups, e.g., connected components~\cite{DBLP:journals/cviu/YeungYL98,zhang2007proteomic,DBLP:conf/wcre/SurianLL10} or cliques~\cite{DBLP:conf/ercimdl/TsatsaronisVTRNSZ11,10.1371/journal.pcbi.1000108,li2005interaction}. We refer to this approach as \emph{rule-based} clustering.

Second, applications can use community detection algorithms. Although there are many conflicting definitions of communities, a community is typically characterized by a large number of internal relationships and a small number of relationships with entities outside the community. Applications of community detection include grouping related papers in a citation network~\cite{DBLP:conf/cscw/JacoviSGUSM06}, identifying city regions~\cite{DBLP:journals/tvcg/HuangZMYYZ16}, and finding people with shared interests~\cite{DBLP:journals/cacm/SchwartzW93}. We refer to this approach as \emph{community-based} clustering.

\subsection{Subgraph Isomorphism and Patterns}
\label{sec:analysis:isomorphism}

\emph{Keywords used to identify uses: subgraph, isomorphism, motif, pattern matching, cycle, clique, star, mesh.}

The subgraph isomorphism problem is the task of identifying a subgraph of the input graph that is isomorphic to a second graph (the \emph{pattern}). The outcome of the subgraph isomorphism problem may be used directly by applications, e.g., to lookup chemical compounds by structure in a database~\cite{nijssen2004frequent,wegner2006data}, or to determine the similarity of two program dependence graphs for plagiarism detection~\cite{DBLP:conf/kdd/LiuCHY06}. Subgraph isomorphism is also used as a component for solving other problems. For example, frequent subgraph mining uses subgraph isomorphism to identify subgraphs that occur frequently in a set of graphs. This technique can be used to discover common web browsing patterns~\cite{heydari2009graph}, localize software bugs~\cite{DBLP:conf/pkdd/EichingerBH08}, classify text~\cite{DBLP:conf/sgai/JiangCSZ09}, etc. More recently, discriminative subgraph mining algorithms have been developed to identify discriminative subgraphs, That is, given a set of labeled input graphs, these algorithms identify subgraphs that typically occur in graphs with one label, but not in graphs with another label. An exemplary use case for discriminative subgraph mining is locating bugs through program flow graphs, each labeled as either ``triggered bug'' or ``did not trigger bug''~\cite{DBLP:conf/issta/ChengLZWY09}.

Related to subgraph isomorphism is the problem of identifying subgraphs that match a more loosely defined pattern, e.g., a clique or a star. Unlike the patterns used as input to subgraph isomorphism, stars and cliques describe a general structure of vertices and edges, but no fixed size\footnote{It is possible to identify these specific patterns by creating stars or cliques of al possible sizes (up to the size of the graph) and using these as concrete patterns for a subgraph isomorphism algorithm. However, this process is inefficient and may not be feasible for more complex patterns. Incidentally, this approach proves the NP-completeness of subgraph isomorphism by reduction from the clique problem.}. We refer to this set of problems as \emph{pattern detection}. Example applications include identifying protein complexes as clique patterns~\cite{li2005interaction} and identifying malicious hosts or domain names as star patterns in a DNS failure graph~\cite{DBLP:conf/icnp/JiangCJLZ10}.

\subsection{Graph Mutation}
\label{sec:analysis:mutation}

Many applications add properties to an input graph as part of their graph analysis method, e.g., by adding a distance attributed to every vertex in a shortest path computation, or a component identifier as output of a connected components algorithm. In contrast, the structure of the graph is often static: vertices and edges are neither added nor removed. We identify several use cases for mutating a graph's structure as part of the analysis of a graph. We distinguish between graph construction, i.e., extending the input graph or constructing a new graph algorithmically, and graph reduction, i.e., removing or merging vertices or edges from the input graph. In our taxonomy, we do not consider extracting a subgraph from the input graph to be graph mutation.

Graph construction is used to dynamically generate parts of a planning graph that are relevant to the analysis, especially when the full planning graph contains many irrelevant vertices and may not fit in memory~\cite{DBLP:journals/ai/BlumF97,DBLP:journals/jair/Hong01}.
Graph reduction is often used to merge groups of related vertices into one vertex per group, thereby creating a new, smaller graph that hides relationships \emph{within} groups and exposes relationships \emph{between} groups. This approach is used to group protein complexes in protein interaction networks~\cite{10.1371/journal.pcbi.1000108}, communities of authors in citation graphs~\cite{DBLP:conf/ercimdl/TsatsaronisVTRNSZ11}, city regions in a road network~\cite{DBLP:journals/tvcg/HuangZMYYZ16}, etc.
Other use cases include simulating brain damage models by deleting edges from a brain network~\cite{stam2008graph} and replacing frequently occurring patterns in a graph with a single copy and an associated count to shrink the graph~\cite{DBLP:conf/kdd/MaxwellBR10}.

\subsection{Other}
\label{sec:analysis:other}

We identify three applications whose primary method of graph analysis does not match any of the aforementioned classes of methods. First, Calderara et al.~\cite{DBLP:journals/cviu/CalderaraHPCT11} use spectral graph theory to characterize a graph constructed from paths formed by moving people and to identify anomalous paths. Second, Polonium~\cite{chau2011polonium} uses an algorithm based on belief propagation to identify files that are likely to be malicious. Third, Zhang et al.~\cite{zhang2007proteomic} use a greedy set cover algorithm to identify a feasible set of proteins to explain the observed set of protein fragments.

\section{Taxonomy}
\label{sec:taxonomy}

\newcommand{\Y}{\cmark}
\let\oldO\O
\let\oldS\S
\renewcommand*{\O}{$\cdot$}
\renewcommand*{\S}{$\star$}
\newcommand{\T}{\textasciitilde}
\newcommand{\header}[1]{\sidetext{\parbox{4.2cm}{\textbf{#1}}}}

\begin{table*}[!t]
	\centering
	\caption{Taxonomy of surveyed articles. A dot indicates that a graph feature or analysis method was not identified in an article. Legend by column:
		\textbf{Weighted}: vertex (\textbf{V}), edge (\textbf{E}), edge weights removed after filtering low weight edges from input (\textbf{\S}).
		\textbf{Heterogeneous}: vertex (\textbf{V}), edge (\textbf{E}).
		\textbf{Properties}: vertex (\textbf{V}), edge (\textbf{E}).
		\textbf{Neighborhood Statistics}: clustering coefficient (\textbf{C}), degree distribution (\textbf{D}), other (\textbf{\S}).
		\textbf{Paths \& Traversals}: shortest path (\textbf{S}), average path length (\textbf{A}), maximum path length (\textbf{M}), traversal (\textbf{T}), other (\textbf{\S}).
		\textbf{Connectivity}: weakly connected components (\textbf{W}), strongly connected components (\textbf{S}).
		\textbf{Centrality \& Ranking}: betweenness centrality (\textbf{B}), closeness centrality (\textbf{C}), degree centrality (\textbf{D}), PageRank (\textbf{P}), other (\textbf{\S}).
		\textbf{Clustering}: rule-based (\textbf{R}), community-based (\textbf{C}).
		\textbf{Subgraph Isomorphism \& Mining}: pattern detection (\textbf{P}), subgraph isomorphism (\textbf{S}), frequent subgraph mining (\textbf{F}), discriminative subgraph mining (\textbf{D}).
		\textbf{Graph Mutation}: construction (\textbf{C}), reduction (\textbf{R}).}
	\label{tab:taxonomy:taxonomy}
	\begin{tabular}{cccccccccccccc}
		\toprule
		& & \multicolumn{5}{c}{\textbf{Graph Features}} & \multicolumn{7}{c}{\textbf{Graph Analysis Methods}} \\
		\cmidrule(lr){3-7} \cmidrule(lr){8-14}
		\textbf{Domain} & \textbf{Ref.} & \header{Directed Edges} & \header{Weighted} & \header{Heterogeneous} & \header{Properties} & \header{Temporal} & \header{Neighborhood\\Statistics} & \header{Paths \& Traversals} & \header{Connectivity} & \header{Centrality \& Ranking} & \header{Clustering} & \header{Subgraph Isomorphism\\\& Patterns} & \header{Graph Mutation} \\
		\midrule
		\multirow{8}{*}{Biology}                               & \cite{barberan2012using}                     & \O & \O & \O & \O & \O & C    & A   & \O & \O    & \O & \O & \O \\
		                                                       & \cite{bilgin2010ecm}                         & \O & \O & \O & \O & \O & CD   & AM  & W  & \O    & \O & \O & \O \\
		                                                       & \cite{bilgin2007cell}                        & \O & \O & \O & \O & \O & CD   & AM  & W  & C     & \O & \O & \O \\
		                                                       & \cite{koyuturk2004efficient}                 & \Y & \O & \O & \O & \O & \O   & \O  & \O & \O    & \O & F  & \O \\
		                                                       & \cite{li2005interaction}                     & \O & \O & \O & \O & \O & \O   & \O  & \O & \O    & R  & P  & R  \\
		                                                       & \cite{10.1371/journal.pone.0012912}          & \Y & E  & \O & \O & \O & \O   & T   & S  & \O    & \O & \O & R  \\
		                                                       & \cite{10.1371/journal.pcbi.1000108}          & \O & \O & \O & \O & \O & \O   & \O  & W  & \O    & R  & P  & R  \\
		                                                       & \cite{zhang2007proteomic}                    & \O & \O & V  & \O & \O & \O   & \O  & W  & \O    & R  & \O & R  \\
		\midrule                                                                                                                                                                
		\multirow{8}{*}{Neuroscience}                          & \cite{BARTOLOMEI20062039}                    & \O & \S & \O & \O & \O & C    & A   & \O & \O    & \O & \O & \O \\
		                                                       & \cite{caeyenberghs2012graph}                 & \O & E  & \O & \O & \O & C\S  & A   & \O & B     & \O & \O & \O \\
		                                                       & \cite{fallani2007cortical}                   & \Y & \S & \O & \O & \O & CD\S & A   & \O & \O    & \O & \O & \O \\
		                                                       & \cite{10.1371/journal.pone.0013788}          & \O & \S & \O & \O & \O & C    & A   & \O & \O    & \O & \O & \O \\
		                                                       & \cite{stam2008graph}                         & \O & E  & \O & \O & \O & C    & A   & \O & \O    & \O & \O & R  \\
		                                                       & \cite{supekar2008network}                    & \O & \S & \O & \O & \O & CD   & A   & \O & \O    & \O & \O & \O \\
		                                                       & \cite{wilke2011graph}                        & \Y & \S & \O & \O & \Y & \O   & \O  & \O & B     & \O & \O & \O \\
		                                                       & \cite{zamora2009graph}                       & \Y & \O & \O & \O & \O & CD   & A\S & \O & \O    & C  & \O & \O \\
		\midrule                                                                                                                                                                
		\multirow{7}{*}{Security}                              & \cite{DBLP:journals/cviu/CalderaraHPCT11}    & \Y & E  & \O & \O & \O & \O   & \O  & \O & \O    & \O & \O & \O \\
		                                                       & \cite{chau2011polonium}                      & \O & \O & V  & \O & \O & \O   & \O  & \O & \O    & \O & \O & \O \\
		                                                       & \cite{DBLP:conf/icnp/JiangCJLZ10}            & \O & \O & V  & \O & \Y & \O   & \O  & W  & \O    & C  & P  & \O \\
		                                                       & \cite{DBLP:conf/ccs/KwonMJBD15}              & \Y & \O & \O & VE & \Y & CD\S & M   & \O & \O    & \O & \O & \O \\
		                                                       & \cite{DBLP:conf/uss/NagarajaMHCB10}          & \O & \O & \O & \O & \O & \O   & T   & \O & \O    & C  & \O & \O \\
		                                                       & \cite{DBLP:conf/csiirw/NoelJ14}              & \Y & \O & \O & \O & \O & \O   & M   & WS & \O    & \O & \O & \O \\
		                                                       & \cite{DBLP:conf/nspw/PhillipsS98}            & \Y & E  & \O & VE & \O & \O   & S\S & \O & \O    & \O & \O & \O \\
		\midrule                                                                                                                                                                
		\multirow{7}{*}{\shortstack[1]{Software\\Engineering}} & \cite{DBLP:conf/issta/ChengLZWY09}           & \Y & \O & E  & V  & \O & \O   & \O  & \O & \O    & \O & D  & \O \\
		                                                       & \cite{DBLP:conf/pkdd/EichingerBH08}          & \Y & E  & \O & V  & \O & \O   & T   & \O & \O    & \O & F  & \O \\
		                                                       & \cite{DBLP:conf/usenix/GuoZLYLDLZ11}         & \Y & \O & E  & VE & \Y & \O   & T   & \O & \O    & \O & \O & R  \\
		                                                       & \cite{DBLP:conf/kdd/LiuCHY06}                & \Y & \O & E  & \O & \O & \O   & \O  & \O & \O    & \O & S  & \O \\
		                                                       & \cite{DBLP:conf/kdd/MaxwellBR10}             & \Y & \O & \O & VE & \O & \O   & \O  & \O & \O    & \O & F  & R  \\
		                                                       & \cite{DBLP:conf/wcre/SurianLL10}             & \O & \O & \O & \O & \O & D    & \O  & W  & \O    & R  & SF & \O \\
		                                                       & \cite{DBLP:conf/icse/ZimmermannN08}          & \Y & \O & \O & \O & \O & C\S  & \O  & W  & BCD\S & \O & P  & \O \\
		\bottomrule
	\end{tabular}
\end{table*}

\begin{table*}[!t]
	\centering
	\caption*{Table \ref{tab:taxonomy:taxonomy}: Continued.}
	\begin{tabular}{cccccccccccccc}
		\toprule
		& & \multicolumn{5}{c}{\textbf{Graph Features}} & \multicolumn{7}{c}{\textbf{Graph Analysis Methods}} \\
		\cmidrule(lr){3-7} \cmidrule(lr){8-14}
		\textbf{Domain} & \textbf{Ref.} & \header{Directed Edges} & \header{Weighted} & \header{Heterogeneous} & \header{Properties} & \header{Temporal} & \header{Neighborhood\\Statistics} & \header{Paths \& Traversals} & \header{Connectivity} & \header{Centrality \& Ranking} & \header{Clustering} & \header{Subgraph Isomorphism\\\& Patterns} & \header{Graph Mutation} \\
		\midrule
		\multirow{7}{*}{\shortstack[1]{Logistics \&\\Planning}} & \cite{DBLP:journals/ai/BlumF97}              & \Y & \O & V  & \O & \O & \O   & T   & \O & \O    & \O & \O & C  \\
		                                                        & \cite{DBLP:conf/mobihoc/DalyH07}             & \O & \O & \O & \O & \O & \O   & \O  & \O & B     & \O & \O & \O \\
		                                                        & \cite{DBLP:conf/aips/Helmert04}              & \Y & \O & \O & \O & \O & \O   & T   & \O & \O    & \O & \O & \O \\
		                                                        & \cite{DBLP:journals/jair/Hong01}             & \Y & \O & V  & \O & \O & \O   & T   & \O & \O    & \O & \O & C  \\
		                                                        & \cite{DBLP:journals/tvcg/HuangZMYYZ16}       & \Y & V  & \O & \O & \Y & \O   & \O  & \O & BP    & C  & \O & R  \\
		                                                        & \cite{porta2006networkDual}                  & \O & \O & \O & \O & \O & CD\S & A   & \O & \O    & \O & \O & \O \\
		                                                        & \cite{porta2006networkPrimal}                & \O & E  & \O & \O & \O & \O   & \O  & \O & BCD\S & \O & \O & \O \\
		\midrule                                                                                                                                                                 
		\multirow{4}{*}{Sociology}                              & \cite{DBLP:conf/socialcom/KhrabrovC10}       & \Y & \O & \O & \O & \Y & \O   & \O  & \O & P\S   & \O & \O & \O \\
		                                                        & \cite{DBLP:journals/cacm/SchwartzW93}        & \Y & \O & \O & \O & \O & \O   & AM  & W  & \O    & \O & \O & \O \\
		                                                        & \cite{DBLP:conf/wise/YamaguchiTAK10}         & \Y & E  & VE & \O & \O & \O   & \O  & \O & P\S   & \O & \O & \O \\
		                                                        & \cite{DBLP:conf/sigir/YangLLR12}             & \Y & \O & V  & \O & \O & \O   & \O  & \O & \S    & \O & \O & \O \\
		\midrule                                                                                                                                                                 
		\multirow{3}{*}{Psychology}                             & \cite{bertola2014graph}                      & \Y & \O & \O & \O & \O & C\S  & A   & S  & \O    & \O & \O & \O \\
		                                                        & \cite{lerner2009network}                     & \O & \O & \O & \O & \O & C    & AM  & \O & \O    & \O & \O & \O \\
		                                                        & \cite{mota2014graph}                         & \Y & \O & \O & \O & \O & C\S  & A   & WS & \O    & \O & \O & \O \\
		\midrule                                                                                                                                                                 
		\multirow{3}{*}{Science}                                & \cite{DBLP:journals/socnet/Gondal11}         & \O & \O & \O & V  & \O & D    & A   & \O & \O    & \O & \O & R  \\
		                                                        & \cite{DBLP:conf/cscw/JacoviSGUSM06}          & \Y & \O & \O & \O & \Y & \O   & \O  & \O & B     & C  & \O & \O \\
		                                                        & \cite{DBLP:conf/ercimdl/TsatsaronisVTRNSZ11} & \O & E  & \O & \O & \Y & \O   & \O  & \O & \O    & R  & P  & R  \\
		\midrule                                                                                                                                                                 
		\multirow{2}{*}{Chemistry}                              & \cite{nijssen2004frequent}                   & \O & \O & \O & VE & \O & \O   & \O  & \O & \O    & \O & S  & \O \\
		                                                        & \cite{wegner2006data}                        & \O & \O & \O & VE & \O & \O   & \O  & \O & \O    & \O & S  & \O \\
		\midrule                                                                                                                                                                 
		\multirow{2}{*}{Finance}                                & \cite{iori2008network}                       & \O & E  & \O & \O & \Y & CD\S & M   & \O & \O    & \O & \O & \O \\
		                                                        & \cite{wang2012visibility}                    & \O & \O & \O & \O & \O & CD\S & A   & \O & D     & C  & \O & \O \\
		\midrule                                                                                                                                                                 
		\multirow{2}{*}{Linguistics}                            & \cite{biemann2006chinese}                    & \O & E  & \O & \O & \O & \O   & \O  & \O & \O    & C  & \O & \O \\
		                                                        & \cite{DBLP:conf/sgai/JiangCSZ09}             & \Y & VE & VE & \O & \O & \O   & \O  & \O & \O    & \O & F  & \O \\
		\midrule                                                                                                                                                                 
		\multirow{7}{*}{Other}                                  & \cite{heydari2009graph}                      & \Y & V  & \O & \O & \O & \O   & \O  & \O & \O    & \O & F  & \O \\
		                                                        & \cite{jiang2013visibility}                   & \O & \O & \O & \O & \O & D    & \O  & \O & \O    & \O & \O & \O \\
		                                                        & \cite{LEE2003103}                            & \O & \O & V  & \O & \O & \O   & \O  & \O & \O    & \O & S  & R  \\
		                                                        & \cite{DBLP:journals/jiis/MirzaKR03}          & \Y & \O & V  & \O & \O & \O   & A   & W  & \O    & \O & \O & \O \\
		                                                        & \cite{DBLP:journals/corr/abs-0903-0194}      & \Y & \O & \O & \O & \O & D\S  & AM  & WS & P     & C  & \O & \O \\
		                                                        & \cite{0295-5075-97-5-50002}                  & \O & \O & \O & \O & \O & D    & \O  & \O & \O    & \O & \O & \O \\
		                                                        & \cite{DBLP:journals/cviu/YeungYL98}          & \Y & \O & \O & \O & \O & \O   & \O  & S  & \O    & R  & \O & \O \\
		\bottomrule
	\end{tabular}
\end{table*}

\renewcommand*{\O}{\oldO}
\renewcommand*{\S}{\oldS}
\begin{table*}[!t]
	\centering
	\caption{Summary of the number of applications that use a given graph feature or analysis method, derived from the taxonomy presented in Table~\ref{tab:taxonomy:taxonomy}.}
	\label{tab:taxonomy:summary}
	\begin{tabular}{lcl}
		\toprule
		                                                                                 & \textbf{Total \# Applications} & \textbf{\# Applications Per Option} \\ \midrule
		\textbf{Graph Features}                                                          &                                &  \\ \midrule
		\quad Directed Edges                                                             &               31               & \cmark: 31                          \\
		\quad Weighted                                                                   &               19               & V: 3, E: 12, $\star$: 5             \\
		\quad Heterogeneous                                                              &               13               & V: 10, E: 5                         \\
		\quad Properties                                                                 &               9                & V: 9, E: 6                          \\
		\quad Temporal                                                                   &               9                & \cmark: 9                           \\ \midrule
		\textbf{Graph Analysis Methods}                                                  &                                &  \\ \midrule
		\quad Neighborhood Statistics                                                    &               23               & C: 18, D: 14, $\star$:10            \\
		\quad Paths \& Traversals                                                        &               30               & S: 1, A: 19, M:8, T: 7, $\star$: 2  \\
		\quad Connectivity                                                               &               15               & W: 12, S: 6                         \\
		\quad Centrality \& Ranking                                                      &               13               & B: 7, C: 3, D: 3, P: 4, $\star$: 5  \\
		\quad Clustering                                                                 &               14               & R: 6, C: 8                          \\
		\quad Subgraph Isomorphism \& Patterns                                           &               16               & P: 5, S: 5, F: 6, D: 1              \\
		\quad Graph Mutation                                                             &               13               & R: 11, C: 2                         \\ \bottomrule
	\end{tabular}
\end{table*}

As described in Section~\ref{sec:method:analysis}, we classified for every selected application the graph features and analysis methods they use. The results of this process are presented as a taxonomy in Table~\ref{tab:taxonomy:taxonomy}. The table depicts for each class of graph features (see Section~\ref{sec:graph}) and analysis methods (see Section~\ref{sec:analysis}, except ``other'' methods) whether it is used by a given application, and, where applicable, which variants of a feature or method are used. The variants are listed in the caption of Table~\ref{tab:taxonomy:taxonomy} and described in more detail in their respective subsections in Sections~\ref{sec:graph} and~\ref{sec:analysis}. Table~\ref{tab:taxonomy:summary} summarizes the frequency of occurrence for each class and each variant of graph features and analysis methods.

Overall, we find significant diversity in the graphs and analysis methods used by the surveyed applications. As depicted in Table~\ref{tab:taxonomy:summary}, all classes of graph features are present in at least 9 applications (15\%), whereas all classes of analysis methods are present in at least 13 applications (\mytilde{}22\%). Variations of each class tend to differ significantly in the number of occurrences, e.g., of applications using paths and traversals, all-pair shortest path algorithms are used by 19 applications, whereas only 1 application uses single-source shortest paths.

As shown in Table~\ref{tab:taxonomy:taxonomy}, there is not only significant diversity in the graph features and analysis methods used, but also in the \emph{combinations of} features and methods. Some combinations occur with high frequency, e.g., local clustering coefficient and all-pair shortest paths are combined in many applications to identify ``small-world'' properties in a graph. Many other combinations are infrequent, but there is no clear segregation between classes of features or methods. This suggest that support for all features and methods should be supported in a single graph analysis platform.

We observe that diversity was not found across all application domains. In particular, we find significant overlap in the approaches used within the neuroscience, psychology, and chemistry domains, although the number of surveyed articles in each domain is too small to make sound claims about the representativeness of our results for individual domains.

Finally, we note that two entries in the taxonomy have no associated analysis methods: applications~\cite{DBLP:journals/cviu/CalderaraHPCT11} and~\cite{chau2011polonium} in the security domain. As described in Section~\ref{sec:analysis:other}, these applications use graph analysis methods that do not fit any of the identified classes and are not used by any other application.

\section{Future Research Directions}
\label{sec:future}

Using the analysis of our taxonomy as presented in Section~\ref{sec:taxonomy}, we present several directions for future research in the development and tuning of graph analysis systems with a focus on functionality and performance. First, the significant diversity in graph features, analysis methods, and their combinations used in real-world applications suggests a need for graph analysis platforms with support for many classes of graphs and analysis methods. Recent developments of graph analysis platforms have focused on either graph databases (e.g., Neo4j~\cite{neo4j-web}) or (parallel) graph processing platforms (Doekemeijer et al.~\cite{doekemeijer2014survey} surveyed many). Graph databases have extensive support for mutable, heterogeneous property graphs, for path queries, and for pattern matching. In contrast, graph processing platforms typically focus on static graphs and large-scale analytics using methods such as centrality and clustering. We do observe a trend towards bridging this gap, e.g., with Neo4j announcing support for graph algorithms in their database~\cite{neo4j-algorithms}.

Second, many applications use more than one method of analysis on the same graph, so there is a need for platforms that allow either interactive analysis or workflows of algorithms for batch processing. Typical graph databases already offer this functionality by providing a service that can be interactively queried, but graph processing platforms typically require loading a copy of the graph from disk for each job (i.e., a single algorithm). Future research on graph processing platforms may study the reuse of a graph stored in memory to speed up sequences of jobs, or even multiple algorithms executing in parallel on the same graph. This research, as well as the development of new benchmarks to assess support for graph analysis workflows, may require additional insight in how practitioners use graph analysis platforms (e.g., do they define in advance a set of algorithms to run, or is the selection of analysis methods guided by results of the previous method?)

Third, we observe that many graphs analyzed in practice share a common set of high-level features, as evident from our taxonomy, but that the structural properties of graphs vary widely and are known to impact significantly the performance of graph analysis~\cite{iosup2016ldbc}. Although some structural characteristics are well-defined, e.g., bipartite graphs, there is not comprehensive set of metrics that captures all structural properties relevant to graph analysis performance. Ongoing work on this topic includes the development of synthetic dataset generators that aim to replicate various structural properties of real-world datasets, e.g., DataSynth~\cite{prat2017towards} and Darwini~\cite{edunov2016darwini}.

Finally, we find that many surveyed applications operate on small graphs, i.e., under a million vertices and edges\footnote{Some surveyed articles do not explicitly state the size of their graph, or they present a general solution without a concrete input graph. We have not quantified the precise scales of each application's graph, but draw conclusions based on estimations.}. Most of the larger graphs, i.e., hundreds of millions of edges and more, were found in domains studying digital networks, including social media networks, software engineering networks, and computer security networks. The largest graph we identified is used to identify malware and consists of nearly 1 billion vertices and 37 billion edges~\cite{chau2011polonium}. These findings contradict a recent user study by Sahu et al.~\cite{sahu2017ubiquity} which found that large-scale graphs are more frequent and can be found across many domains. Possible explanations for this contradiction include the more recent data used by the user study\footnote{The development of large-scale graph processing platforms did not gain significant traction until the publication of Pregel in 2010~\cite{doekemeijer2014survey}, so we believe it is feasible that adoption of large-scale graph analysis techniques has increased between the publication of our surveyed articles (many of which appeared before 2010) and the 2017 study by Sahu et al.}, a potential bias in~\cite{sahu2017ubiquity} due to surveying users of specific software products (of which many explicitly target large-scale graph analysis), and a disparity between graphs used in academia versus industry.

\section{Reproducibility}

To facilitate the validation and reproduction of our results, we provide the data collected and produced while performing this survey as open-access data~\cite{hegeman_tim_2018_1298640}. We provide a complete list of search results obtained via Google Scholar, a shortlist of relevant identified results (see Section~\ref{sec:method:identification}), and the resulting analysis and characterization (see Sections~\ref{sec:method:analysis},~\ref{sec:taxonomy}).
\section{Conclusion}

The development and tuning of graph analysis platforms benefits greatly from knowledge of their practical applications. However, such knowledge requires a broad view of graph analysis use cases across many domains and an in-depth view of the datasets and algorithms they use.
Earlier work focuses on few applications, provides a high-level view of many applications, or studies in depth a small set of algorithms.

In this work we made a threefold contribution towards gaining insight in the practical use of graph analysis:
\begin{enumerate*}[label=(\roman*)]
	\item we defined and applied a systematic method for identifying and selecting relevant literature on graph analysis applications across many domains,
	\item we presented a taxonomy of the graph features and graph analysis methods used in practice, and
	\item we proposed several directions for future research in developing and tuning graph analysis platforms.
\end{enumerate*}

Our primary observation is the large diversity in the domains to which the identified applications belong and in the classes of graph features and analysis methods they use. From 5 classes of graph features and 7 classes of analysis methods, each was encountered in at least 15\% of surveyed applications.
We further observe that most applications use multiple analysis methods and that many classes of analysis are combined in practice.
In contrast, some domains show significant homogeneity, especially neuroscience, a domain that focuses on the analysis of brain networks and has seemingly developed a common set of techniques for analyzing such networks.

We conclude that future research in the development and tuning of graph analysis platforms should focus on integrating support for a wide range of graph features and analysis methods, in contrast to the dichotomy of graph databases and parallel graph processing frameworks that currently exists. Additionally, the combination of multiple analysis methods, interactively or as a workflow, should be further investigated and supported in future systems. Finally, to understand the differences in graphs across domains and their impact on performance, additional research is needed to characterize the structural characteristics of graphs beyond simple structures such as bipartite graphs.

\bibliography{references-surveyed-articles,references-other}

\newpage
\appendix

\end{document}